# The importance of thermal gradients on the vortex dynamics and magnetic behavior of mesoscopic superconducting samples

E C S Duarte[1], A Presotto[1], D Okimoto[1], V S Souto[1], E Sardella[2] and R Zadorosny[1]

[1]Departamento de Física e Química, Faculdade de Engenharia de Ilha Solteira, UNESP - Univ Estadual Paulista, Caixa Postal 31, 15385-000 Ilha Solteira-SP, Brazil
[2]Departamento de Física, Faculdade de Ciências, UNESP - Univ Estadual Paulista, Caixa Postal 473, 17033-360, Bauru-SP, Brazil
E-mail: rafael.zadorosny@unesp.br

**Abstract.**
Usually, the measurements of electronic and magnetic properties of superconducting samples are carried out under a constant temperature bath. On the other hand, thermal gradients induce local variation of the superconducting order parameter, and the vortex dynamics can present interesting behaviors. In this work, we solved the time-dependent Ginzburg-Landau equations simulating samples under two different thermal gradients, and considering two values of the Ginzburg-Landau parameter, $\kappa$. We find out that both parameters, i.e., $\kappa$ and thermal gradients, play an important role on the vortex dynamics and on the magnetization behavior of the samples.

## 1. Introduction

Several applications of superconducting materials lie on the increasing of the critical current density, $J_c$. on the other hand, as $J_c$ is related to the vortex dynamics, the understanding of such issue is of great relevance, and several works have demonstrated that artificial pinning centers disposed in periodic [1, 2, 3, 4, 5, 6] and non-periodic arrays [7, 8, 9] are effective to immobilize the vortices, avoiding their viscous motion. In Ref. [10], the authors studied the pinning properties of a graded distribution of holes in comparison with a regular array of similar holes. It was shown that for low fields, the graded sample presents larger $J_c$ than that one with a regular distribution of holes. Alternatively, the control of the vortex dynamics was also studied in devices for vortex ratchet [11, 12, 13] for which $J_c$ can be controlled by applied ac currents or fields.

Nevertheless, huge vortex penetration, such as flux avalanches phenomenon should be avoided [15] or reflected by a metallic layer [16]. Brisbois et al. [17] investigated the flux penetration in two Nb films with different thickness in its central region where, in one case, was considered a blind hole and in the other one an elevated superconducting region. Additionally, it was also used the Ginzburg-Landau formalism to demonstrate the vortex dynamics of the systems. As expected, in the first configurations, the thinner region acts as a pinning center. However, the configuration with a thicker central portion behaves as a barrier to the entrance of vortices, which preferentially lie on the edges of such region.

Another interesting feature was described by Gladilin and co-workers [18] where the time-dependent Ginzburg-Landau (TDGL) theory was used to study a wedge of a type I superconductor. The system show penetrations of individual vortices from its thinner part which gradually moves to thicker regions. As the dynamics evolves, the vortices begin to coalesce into giant vortices with vorticities of $2^n$ (n = 1; 2).

As a general remark, asymmetric and graded distribution of holes and a non-uniformity of the superconducting region affect both the vortex dynamics and the critical current density of the sample. Besides that, a non-uniform distribution of the superconducting order parameter can also be induced by thermal gradients. Previous works in this issue treated with the temperature dependence of the viscous coefficient [19], thermoelectric phenomena [20, 21, 22] and charge imbalance [23]. On the other hand, it was shown in Ref. [24] that giant vortex states are induced by thermal gradients in such a way that they can be avoided.

Then, in this work, we studied the vortex configurations guided by the field-dependent magnetization of samples with different thermal gradients and $\kappa$ parameters. It was verified that the vortex dynamics and the appearance of a non-conventional magnetization behavior can be controlled by changing the local temperature of the superconducting samples.

## 2. Theoretical Formalism

In 1966, Schmid [25] presented a phenomenological treatment for non-equilibrium superconductivity based on the Ginzburg-Landau theory. It is usually known as TDGL equations. In those equations it is considered the temporal evolution of the superconducting order parameter, $\psi$, and of the vector potential, **A**. In dimensionless units, the TDGL equations are given by

$$u\left(\frac{\partial}{\partial t} + i\phi\right)\psi = -(i\mathbf{\nabla} - \mathbf{A})^2\psi + (1 - T(x) - |\psi|^2)\psi,$$
$$\left(\frac{\partial \mathbf{A}}{\partial t} + \mathbf{\nabla}\phi\right) = J_s - \kappa^2 \mathbf{\nabla} \times \mathbf{\nabla} \times \mathbf{A},$$

with the superconducting current density given by
$$J_s = Re(\psi^*(-i\mathbf{\nabla} - \mathbf{A})\psi).$$

To take into account a thermal gradient, we assume a linear dependence of $T$ along the $x$ axis,

$$T(x) = T_l + \frac{(T_r - T_l)x}{L_x},$$

where $L_x$ is the length of the superconductor along the x direction.

In eq. (4) $T_l$ and $T_r$ are the temperature on the left and right sides of the system, respectively. For all the simulated systems, we assume that the temperature increased from the left to the right side. Here, the distances are measured in units of the coherence length $\xi(0)$ at zero temperature, the magnetic field in units of the bulk upper critical field, $H_{c2}(0)$, and the temperature in units of the critical temperature, $T_c$. The parameter $u$ is the ratio between the relaxation time of $\psi$ and the superconducting characteristic time. We choose $u = 1$ which is accepted in the literature to study vortex dynamics. We used the Coulomb gauge for which the scalar potential $\phi = 0$ for all times and positions. We numerically solved the TDGL equations by the link variable method [26], which preserves the gauge invariance of the equations when they are discretized.

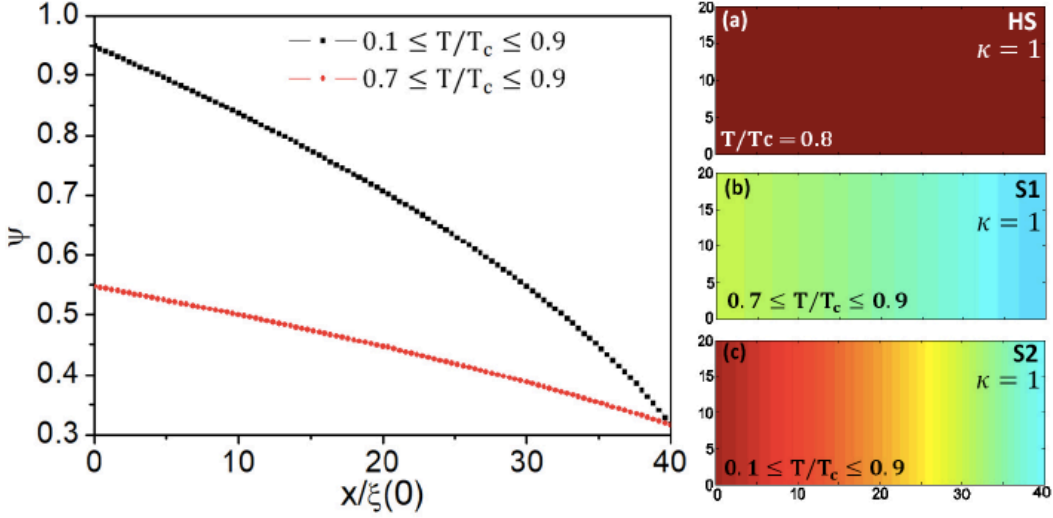

**Figure 1.** The main panel shows the variation of $|\psi|$ along the samples S1, for which $0.7 \leq T/T_c \leq 0.9$, and S2, where $0.1 \leq T/T_c \leq 0.9$. The panels from (a) to (c) present the map of $|\psi|$ for the samples HS at $T = 0.8T_c$, S1, and S2 respectively, and with $\kappa = 1$.

## 3. Results and Discussion

We simulated rectangular samples of dimensions $40\xi(0) \times 20\xi(0)$. The discretization of eqs. (1-4) was made in an uniform mesh grid of eight points per one coherence length, that is, $\Delta x = \Delta = y = 0.125\xi(0)$. Two different values for the coldest and hottest edges were considered, as shown in Figure 1. The samples were labeled as S1 for $0.7\,T_c < T < 0.9\,T_c$ and S2 for $0.1\,T_c < T < 0.9\,T_c$. For the homogeneous samples, HS, the temperature was fixed at $0.8T_c$. All the systems taken in the Meissner state at $H = 0$, and the external magnetic field was increased in steps of $\Delta H = 10^{-3}H_{c2}(0)$ until superconductivity is destroyed. We chose the Ginzburg-Landau parameters $\kappa = 1$ and $\kappa = 5$, which correspond to low and moderate $\kappa$ materials, respectively. Figure 1 shows the behavior of $|\psi|$ through the samples HS, S1 and S2 for $\kappa = 1$ (panels (a), (b) and (c), respectively). In the mixed state, the different samples present distinct response functions (the magnetization curve $M(H)$) as is shown in Figure 2. In samples $S1_{\kappa=5}$ and $S1_{\kappa=1}$, panels (a) and (b) respectively, the $M(H)$ curves are similar to those ones of the homogeneous systems, as shown in the respective insets. On the other hand, the $S2_{\kappa=5}$ and $S2_{\kappa=1}$ samples (panels (c) and (d)) exhibit distinct behaviors when compared with their respective homogeneous samples. To see this difference, firstly, we note that the intensity of the magnetization signal is at least twice as much than that one presented by the similar homogeneous specimens (compare insets of panels (a) and (b) with panels (c) and (d), respectively). This means that the hotter portions work as pinning centers, maintaining the vortices trapped in such regions, which allow the superconductivity to survive in a large portion of the sample. Secondly, the magnetization response of $S2_{\kappa=5}$ samples are very distinct as can be noticed in Figure 3. Figure 3 shows a zoom of the $M(H)$ curve of sample $S2_{\kappa=5}$. For this sample, one can distinguish two steps where the magnetization oscillates around a medium value. For each step we indicate the fields for approximately the same magnetization value and the respective maps of $|\psi|$ are shown in Figure 4. As the vorticity increases along each step, it can be noticed by the color distribution that the left-hand side of the sample remains unaltered, and the number of vortices in this coldest region are lower than that one in the hottest part. This explains the nearly unchanged values of the magnetization. Notice also that from one step to another it occurs a change in the number of linear chains of vortices. For the first step, we have a single chain, whereas for the second step there is two chains. In summary, within each step we have a change in the vorticity, whereas from step to step we see changes of linear chains. The step-chain characteristic of the $M(H)$ curve is very similar to what occurs in a superconducting film in the presence of a parallel external applied magnetic field [27]. Although the sample $S2_{\kappa=5}$ shows this step-behavior in the magnetization, the intensity of the signal always decreases with the vortex penetration.

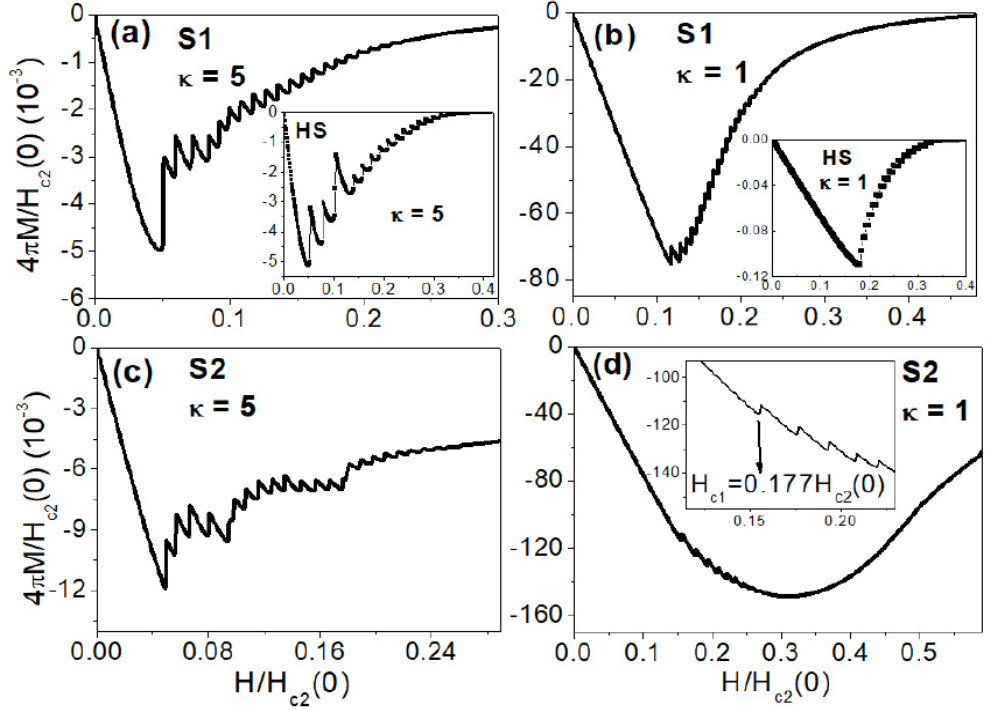

Figure 2. $M(H)$ curves of the samples (a) $S1_{\kappa=5}$, (b) $S1_{\kappa=1}$, (c) $S2_{\kappa=5}$ and (d) $S2_{\kappa=1}$. The homogeneous samples for $\kappa=5$ and $\kappa=1$ are shown in the insets of panels (a) and (b) respectively. One can notice that both the thermal gradient and the value of $\kappa$ control the magnetic behavior of the samples.

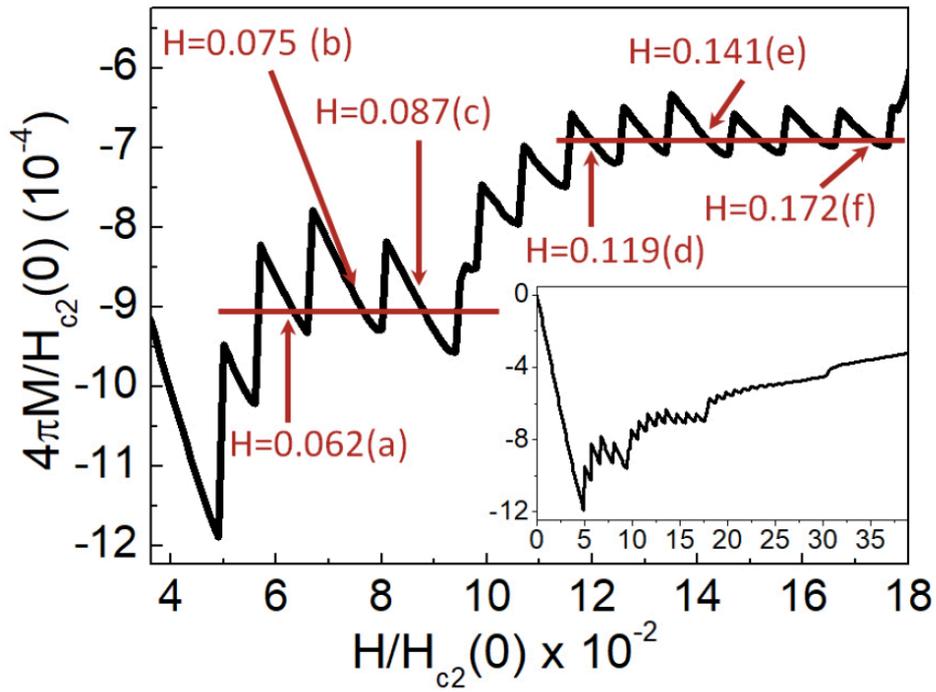

**Figure 3.** Regions where the magnetization have the intensity (solid green lines) for different values of external magnetic field.

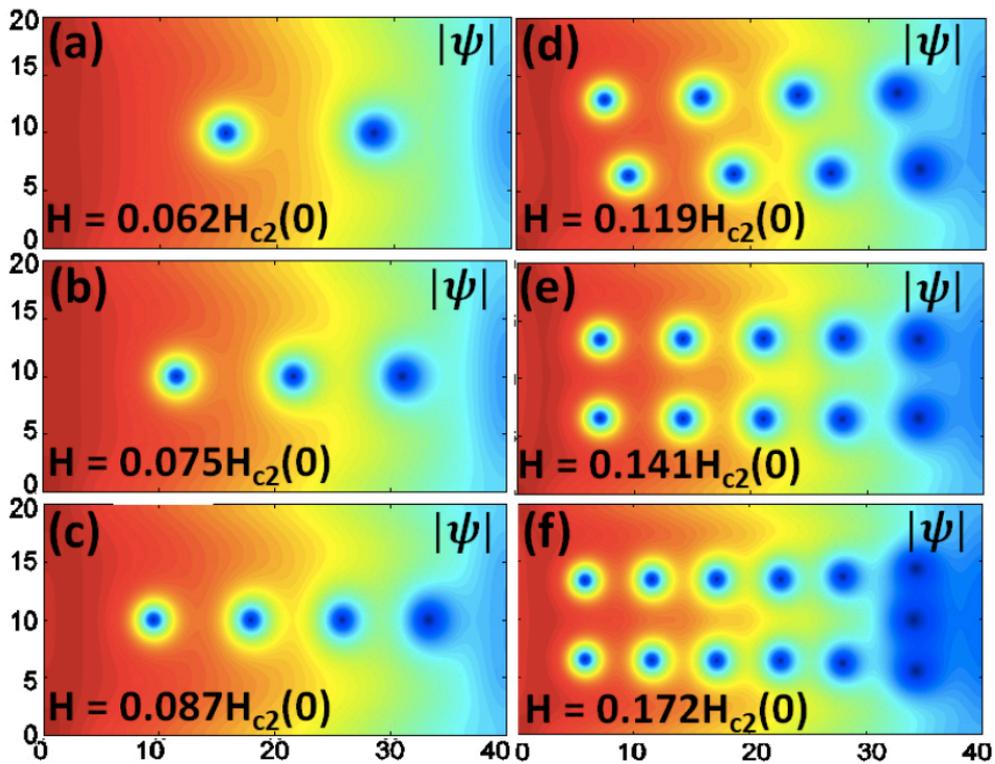

**Figure 4.** Mapping of $|\psi|$ for different values of the external magnetic field indicated on the first region (see Figure 3). For each value of field we can observe the vortices motion and the increase of vorticity.

A quite non-conventional behavior of the $M(H)$ curve also takes place for the sample $S2_{\kappa=1}$, as shown in Figure 2(d). One can notice that besides the first vortex penetrations ($H_{c1} = 0.177 H_{c2}(0)$), the modulus of $M(H)$ remains increasing up to $H_{c1} = 0.307\, H_{c2}(0)$. Such behavior lies on the interaction between the vortices and the shielding currents, $J_{sh}$, which is weaker in the hottest region that behaves as a pinning center.

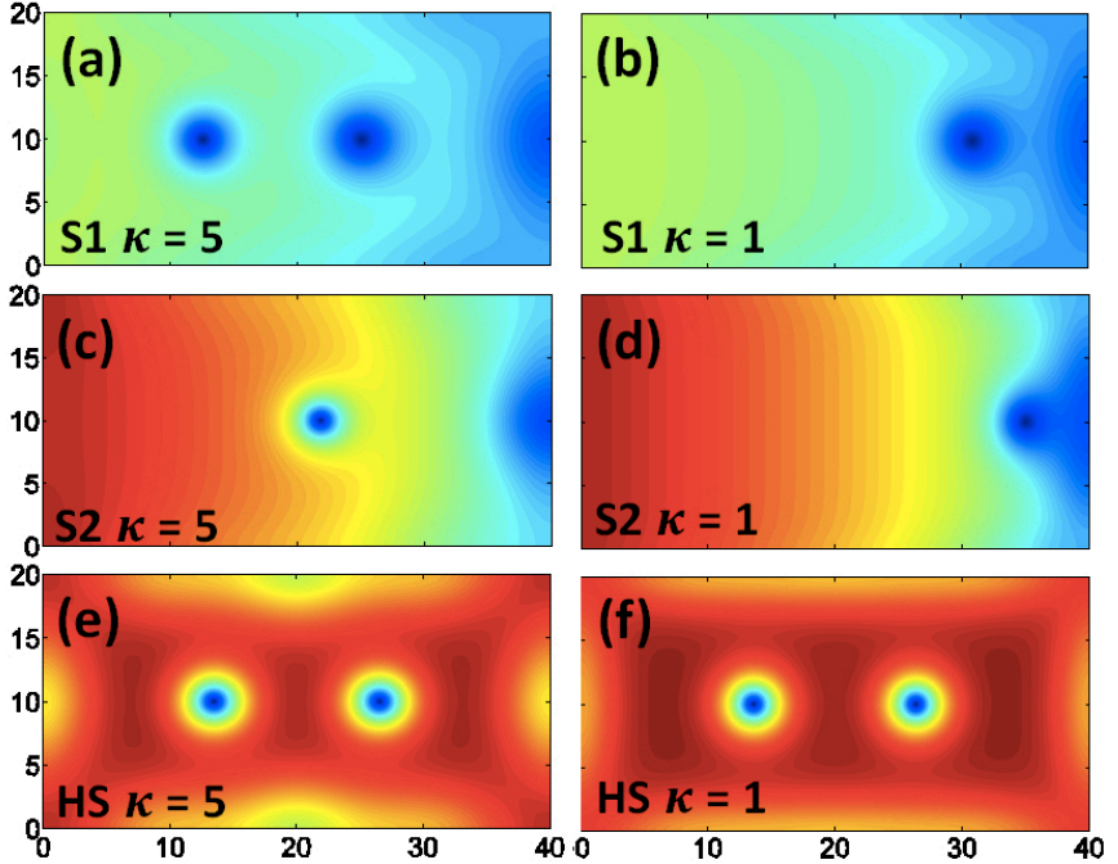

Figure 5. The intensity of $|\psi|$ illustrates the vortex equilibrium configuration in the first vortex penetrations for all simulated systems.

Figure 5 shows the map of $|\psi|$ for the fields indicated in Figure 2(d) where it is evidenced that the hottest region plays the role of a pinning center. The shielding current density, $J_{sh}$, in the coldest region increases by increasing the external magnetic field and the penetrated vortices are trapped in the hottest regions, as shown in Figure 6. In such figure, it is shown a map of $|\psi|$ for the fields indicated in Figure 2(d). The intensity of $J_{sh}$ was also analyzed along the $x$ axis of the $S2_{\kappa=5}$ sample at $y = 0.5\xi(0)$ as is shown in Figure 7. The curves for $H = 0.155, 0.177$ and $0.200 H_{c2}$ present a low increasing of $J_{sh}$ in the coldest region and collapse in a single curve near $x = 32\xi(0)$. Such behavior indicates that the size of the pinning region remains quite unchanged, which means that the vortices do not move through the coldest regions. On the other hand, for $H \geq 0.307 H_{c2}$, the value of $J_{sh}$ increases significantly on the coldest regions, while the vortices move toward such portion of the sample and the modulus of the magnetization decreases smoothly.

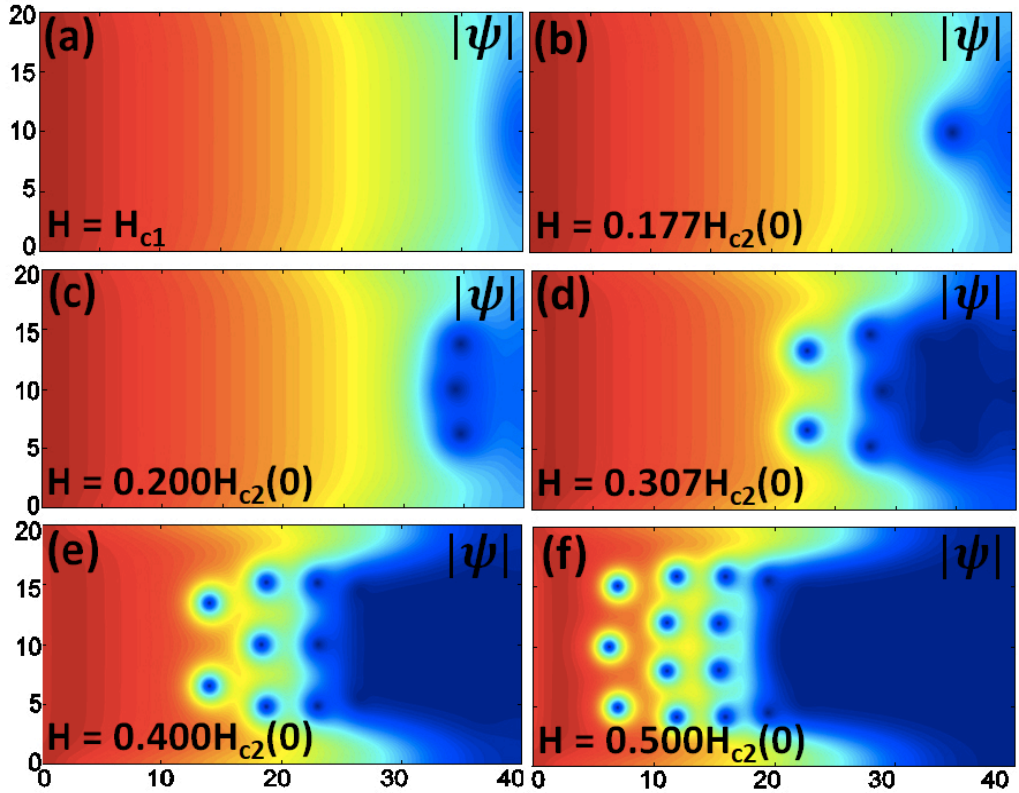

**Figure 6.** The modulus of the order parameter as a function of the external magnetic field for $S2_{\kappa=1}$: (a) $H = H_{c1}$; (b) $H = 0.177H_{c2}(0)$; (c) and $H = 0.200H_{c2}(0)$; (d) $H = 0.307H_{c2}(0)$; (e) $H = 0.400H_{c2}(0)$ (f) $H = 0.500H_{c2}(0)$.

In addition to the non-conventionality of the $S2_{\kappa=1}$ sample, the $M(H)$ curve presents an anomalous response concerning the first vortex penetration. The $H_{c1}$ is lower than the field which corresponds to the local minimum of $M(H)$, (see Figure 2(d)). Such behavior is due to the interaction between the vortices and the shielding currents which is weaker in the hottest region that behaves as a pinning center, as can be observed in Figure 6. As the external applied magnetic field is increased, the current density increases in the coldest region (see Figure 7) which contributes for the increasing of the modulus of $M(H)$, so that the vortices remain in the hottest domain. For fields above $H = 0.307H_{c2}(0)$ the absolute value of $M(H)$ starts decreasing smoothly (see Figure 2(d)). For the $S1_{\kappa=1}$ sample, there is not a great difference between $T_l$ and $T_r$, so the intensity of the shielding current does not change significantly. In such a way that the $M(H)$ curves are similar with those ones for the HS sample (see the inset of Figure 2(a)).

## 4. Conclusion

A non-conventional magnetization response is generated due to a huge thermal gradient in type II mesoscopic superconducting samples for the low $\kappa$ limit ($\kappa = 1$). In this case, the vortices are trapped in the hottest region and then, the shielding currents remains stronger in the coldest one. In other words, the hottest region acts like pinning centers leaving the coldest region as a vortex free zone in the regime $H_{c1} < H \ll H_{c2}$. As a consequence, the modulus of the magnetization continue to increase for $H > H_{c1}$ until reach a great number of vortices, and then begins to exhibit the conventional behavior of mesoscopic systems. Our system has a great similarity with the geometry studied in Ref. [18] which consists of a type I superconductor fabricated as a wedge. The thinner region of the wedge plays the same role as the hottest one in our system. As in the case of wedge variation in type I superconductor studied by Gladilin and co-workers [18], our approach rants the flux motion just in one direction, i.e., from the hottest to the coldest region. Then, this is an important remark for future fluxonic applications.

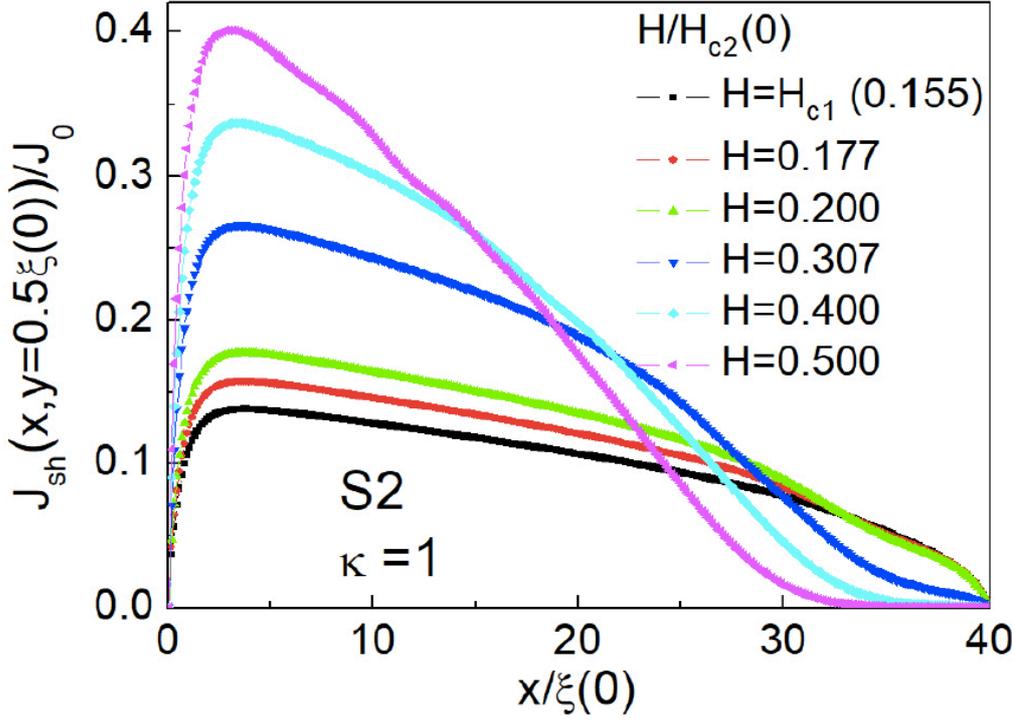

Figure 7. Variation of $|J|$ along the $x$ axis at fixed $y = 0.5\xi(0)$ of S2$_{\kappa=1}$ sample for values of fields indicated in Figure 6.

**Acknowledgements**

We thank the Brazilian Agencies São Paulo Research Foundation, FAPESP, grant 2016/12390-6, and Coordenação de Aperfeiçoamento de Pessoal de Nível Superior - Brasil (CAPES) - Finance Code 001.

ECSD, AP, DO and VSS carried out the numerical simulations and analyzed the data under the supervision of RZ and ES. ES and ECSD wrote the numerical code. RZ conceptualized the work. All authors contributed to the writing of the manuscript.